\documentstyle[12pt]{article}
\begin{document}
\baselineskip 15pt

\title{Quantum superpositions and definite perceptions: envisaging new feasible
experimental tests.}
\author{GianCarlo Ghirardi\footnote{e-mail: ghirardi@ts.infn.it} \\ {\small
Department of
Theoretical Physics of the University of Trieste, and}\\ {\small the Abdus
Salam
International Centre for Theoretical Physics, Trieste, Italy.}}
\date{}
\maketitle

\begin{abstract}

We call attention on the fact that recent unprecedented technological
achievements,
in particular in the field of quantum optics, seem to open the way to new
experimental
tests which might be relevant both for the foundational problems of quantum
mechanics
as well as for investigating the perceptual processes.
\end{abstract}

\section{Introductory considerations.}

	The crucial problem of quantum mechanics can be summarized in a
very elementary way: if
one assumes that the linear evolution law of the theory governes all
natural processes, then, while
in actual measurement processes (and \cite{ghi1} in all {\it those
measurement like processes we are obliged to admit ... are going on more or
less all the time, more or less everywhere}) linear superpositions of
macroscopically different situations occur, we perceive only one among the
many potential
outcomes. In a way or another (more about this in what follows) we have to
recognize that, at least
at the level of our definite perceptions, the linear laws of quantum
mechanics are violated. Here we
will not discuss the extremely delicate and controversial problem of the
perceptual process, but we
will limit ourselves to call attention on the fact that recent
unprecedented technical improvements,
specifically in the field of quantum optics, seem to allow the preparation
of linear superpositions of
states such that the terms of the superposition are able to trigger
definite and different perceptions
of a conscious observer. In spite of the fact that everybody would feel
confident in anticipating the
main characteristics of the final exit of the experiments we are going to
discuss we think that it is
interesting to analyze them since they exhibit some completely new features
and they could, in
principle, allow to get some interesting indications about the percetual
process.

\section{The problem and some standard solutions.}

	Usually one circumvents the just mentioned problem of reconciling
our definite perceptions
with the waviness, the indeterminacy characterizing linear superpositions,
by pointing out that in
actual situations one can avoid to even mention the conscious observer
since the different apparatus
states whose superpositions are generated in measurement like process are
macroscopically
different (the most elementary case being the one of different positions of
the macroscopic pointer
of the apparatus). Even though the issue under discussion is rather
controversial, everybody would
agree that the above remark is absolutely pertinent, at least for all
practical purposes. For clarity
sake, let us elucidate this point by analyzing the puzzling situation we
have just mentioned from the
point of view of various positions which one can take about the theory and
its interpretation.

\begin{description}
\item[The textbook solution.]
With the above expression we denote the position \cite{ghi2, ghi3} of the
founding fathers of the theory,
i.e. the one which is usually referred to as ``the orthodox
interpretation''. It plainly amounts to accept that there are two evolution
laws governing natural processes, the first one characterizing
microscopic systems which is linear and deterministic, the second one
entering into play when
different macroscopic effects are triggered by different microscopic
situations which is
phenomenologically described by the nonlinear and stochastic process of
wave packet reduction.

	Obviously, such a solution has to face problems of consistency and
of lack of mathematical
precision and is considered as unsatisfactory by most scientists involved
in the debate about the
conceptual implications of quantum mechanics, but this is not the point we
want to make here. For
our purposes it is simply useful to remark that if one takes such a
position the problem of our
definite perceptions in the above considered cases simply disappears: even
before we ``look'' at the
apparatus, due to the fact that wave packet reduction has taken place, the
macroscopic system is
already in a macroscopically precise situation which matches our definite
perception about it.

\item[The decoherence argument.]
According to this point of view \cite{ghi4, ghi5} no wave packet reduction
takes place, but the quantum
nature of the ``macroscopic pointer'' (i.e. its being in a superposition)
cannot play, in practice, any
testable role since the interference effects to which it could (in
principle) give rise are hidden or
suppressed by various mechanisms, the most typical one being its
unavoidable coupling with the
environment and the ensuing decoherence. This fact can be put in a precise
mathematical form by
recalling an elementary theorem about composite systems:

Suppose we consider an entangled normalized state $\vert\Psi(1,2)\rangle$
of a composite system
$S = S_{1} + S_{2}$:

\begin{equation}
\vert \Psi(1,2)\rangle\quad =\quad \alpha |\phi^{(1)}\rangle\otimes
|A^{(2)}\rangle +
\beta |\chi^{(1)}\rangle\otimes |B^{(2)}\rangle, \label{ghir1} \end{equation}

and let us evaluate, for such a state, the expectation value of an
arbitrary projection operator
$P_{1}$ of the Hilbert space of $S_{1}$. We have:

\begin{eqnarray}
\langle\Psi(1,2)|P_{1}|\Psi(1,2)\rangle & = &
|\alpha|^{2}\langle\phi^{(1)}|P_{1}|
\phi^{(1)}\rangle + |\beta|^{2}\langle\chi^{(1)}|P_{1}|\chi^{(1)}\rangle
\nonumber \\
& + & 2Re\{\alpha^{*}\beta\langle\phi^{(1)}|P_{1}|\chi^{(1)}\rangle\langle
A^{(2)}|B^{(2)}\rangle\}. \label{ghir2}
\end{eqnarray}

This simple equation expresses the well known fact that if the states
$|A^{(2)}\rangle$ and $|B^{(2)}\rangle$ of the system which is entangled
with $S_{1}$ coincide $(\langle A^{(2)}|B^{(2)}\rangle = 1)$, then the
interference effects are fully exploitable, while if they are orthogonal
$(\langle A^{(2)}|B^{(2)}\rangle = 0)$, any conceivable test on the system
$S_{1}$ gives exactly the same outcome as the one implied by the
statistical mixture of pure
states of $S_{1}$ in which a fraction $|\alpha|^{2}$ of the systems are in
the state
$|\phi^{(1)}\rangle$ and a fraction $|\beta|^{2}$ in the state
$|\chi^{(1)}\rangle$ , i.e. the situation characterized by the statistical
operator:

\begin{equation}
\rho\quad =\quad |\alpha|^{2}|\phi^{(1)}\rangle\langle\phi^{(1)}| +
|\beta|^{2}|\chi^{(1)}\rangle\langle\chi^{(1)}|. \label{ghir3}
\end{equation}

	The conclusions which are relevant for the problem we are
investigating should then be
obvious: the macroscopically different final apparatus states (which must
be identified with the
states $|\phi^{(1)}\rangle$ and $|\chi^{(1)}\rangle$ of the above example)
become immediately entangled with the
environment and the states of the environment associated to the different
terms (i.e. the states
corresponding to $|A^{(2)}\rangle$ and $|B^{(2)}\rangle$) are (practically)
orthogonal. Since when an observer ``looks'' at the pointer he cannot
``perceive'' all environmental degrees of freedom, his perceptions too, can,
in practice, be identified with those triggered by {\it either} the state
$|\phi^{(1)}\rangle$, i.e. the one making legitimate the claim ``the
pointer points at 1'' {\it or} by the state $|\chi^{(1)}\rangle$,
associated to the fact that ``the pointer points at 2''.

\item[The Bohmian point of view.]

	As is well known, the de Broglie-Bohm theory \cite{ghi6, ghi7,
ghi8} 	is a nonlocal and deterministic hidden
variable theory which completely agrees with quantum mechanics and
attributes to all particles of
the universe a perfectly definite position at all times. For the case under
consideration the theory
claims that the pointer is definitely {\it either} in one {\it or} in the
other of the two positions 1 and 2, while its wavefunction (which coincides
with the quantum mechanical one) is different from zero both in
regions 1 and 2.

	Within this framework, the fundamental interplay of the position
variables and the
wavefunction describing the Hilbert space evolution of the physical system
is the fundamental ingredient guaranteeing the
agreement with quantum mechanics. For instance, a particle can be claimed
to follow, in the two
slits experiment, one and only one of the two ``possible paths'' but the
very fact that its
wavefunction is different from zero also on the path which it does not
follow guarantees that, on
repetition of the experiment, one gets the interference pattern on the screen.

	However, it is one of the nicest and most relevant features of the
theory that, while for a
microsystem the full wave function plays the just mentioned fundamental
role, in the case (like the
one which interests us) of a macro-object which is in a precise region but
whose wave function is
different from zero also in another region, one can resort to the
``effective wavefunction'' to describe
its evolution. The effective wavefunction is simply the one obtained by
dropping the so called
``empty wave'', i.e. its part referring to the region where there is no
objetc. This is obviously an
approximation, but it can be proved to hold to an extremely high degree of
accuracy.

	Once more, with reference to our example, we could claim that the
conscious observer is
confronted with an unambiguous and precisely definite macroscopic
situation: the pointer is at a
definite place and only the value of the wavefunction at that place matters.

\item[The spontaneous localization point of view.]

	Recently, a new way out \cite{ghi9, ghi10, ghi11, ghi12} from the
difficulties connected with the measurement problem has been suggested. It
is based on the consideration of nonlinear and stochastic
modifications of the Schr\"{o}dinger equation. Such modifications are due
to terms describing the
occurrence of spontaneous localization processes affecting all elementary
constituents of any
composite system. The characteristic trait of the approach derives from the
fact that the dynamical
modifications are such that they leave (practically) unaltered all
predictions of standard quantum
mechanics for microscopic systems but they induce an extremely rapid
(within millionths of a
second) suppression of the superpositions of macroscopically
distinguishable states.

	With reference to our problem, the various dynamical reduction
models which have been
discussed in the literature lead (even though in a consistent way and as a
consequence of a universal
dynamical equation) to the same conclusion as the standard theory
supplemented by the assumption
of wave packet reduction: the superpositions of macroscopically different
states corresponding to
different pointer positions are dynamically forbidden. Once more, the
conscious observer faces a
macroscopically definite situation: the quantum ambiguity connected with
superpositions has
already been disposed of by the universal dynamics which characterizes such
models.
\end{description}

	The conclusion one can draw from the previous analysis should be
obvious: in a way or
another, quite independently of the position one takes about the theory and its
interpretation\footnote{We do not want to be misunderstood: we consider the
macro-objectification problem as a quite relevant one deserving hard work
by theoretical physicists. But this fact can be disregarded for our present
purposes.}, the standard wisdom about perceptions is restored: the
perceptions of the conscious observer can be considered, {\it for all
practical purposes}, as triggered by unambiguous macroscopic situations. It
is therefore interesting to investigate whether one can devise situations
in which one can be sure that the perceptual process is directly triggered
by a genuine superposition of different states which, when considered by
themselves, would give rise to different perceptions.

	Before coming to discuss this point, a short clarifying digression
is appropriate.

\section{A short digression.}

	To prepare the reader to the subsequent analysis it seems useful to
recall a
question which has been debated in the literature several years ago
\cite{ghi13}.
Suppose we send a spin 1/2 particle in a state which is the eigenstate
corresponding
 to the eigenvalue +1 of the observable $\sigma_{x}$, through a Stern-Gerlach
apparatus devised to measure $\sigma_{z}$. We all know that a subsequent
experiment
 aimed to detect whether the particle is in the upper or in the lower of
the two
possible paths for it will give an unambiguous outcome: {\it either} the
particle
will activate the counter on the upper path {\it or} the one on the lower one.
 Now comes the question: does the reduction of the wave packet take place at
the moment of the detection or has it already taken place during the process
of traversing the inhomogeneous magnetic field? B. d'Espagnat has given the
clear-cut answer: the reduction cannot take place when the particle goes
through
 the region of the Stern-Gerlach apparatus for the simple reason that everybody
 would agree that if one were able to build an apparatus which undoes exactly
what the first one did (i.e. it induces the time-reversed evolution), it would
 recombine the two wave packets and it would restore the spin eigenstate
 $|\sigma_{x} = 1\rangle$ for the particle. As a consequence, a measurement
 of $\sigma_{x}$ would give, with certainty, the outcome +1. On the contrary,
 if consideration were given to the statistical mixture corresponding to the
assumption that the reduction has already taken place, the unfolding of
the ``reverse'' process would assign probability 1/2 to the outcome
 $\sigma_{x} = 1$. The question admits therefore a clear cut experimental
answer.

Obviously, one could object that this is a ``truly gedanken experiment''
and that the test is possible only ``in a science fiction world'' in which
one can actually build an apparatus which undoes precisely what the first
Stern-Gerlach apparatus did to the particle. However, we all know the way
to circumvent this (appropriate) criticism by resorting to systems
involving photons. Such a procedure brings down the above experiment from
``the heaven of the conceptual possibilities'' to ``the real world of the
actual feasibility''.

In fact, let us consider \cite{ghi14} a photon with plane polarization at
45 degrees impinging on a birifrangent crystal whose ordinary ray is
characterized by the horizontal and the extraordinary ray by the vertical
polarization. From the crystal two spatially well separated rays emerge,
or, to be
precise, after having traversed the crystal the photon is in the
superposition of being in the ordinary
ray and having a horizontal polarization and of being in the extraordinary
ray and having a vertical
polarization. To {\bf check} that the superposition {\bf is still present}
is quite easy: one can use a mirror
reflected crystal which recombines the two states and makes them to
interfere. After traversing it
the photon is polarized at 45 degrees, as one can easily check by putting a
polarizer in front of a
photon detector. So we have here an example of a situation in which it is
possible (and easy) to test
that {\it before impinging} on the second crystal the superposition of the
``spatially separated states'' is still there.

	We are now ready to discuss the experiment we want to propose.

\section{A challenge for experimental quantum optics.}

	The idea we are going to present in this Section is based on the
simple remark that the
human visual percetual apparatus is characterized by an extreme
sensitivity. As is well known, the
threshold for visual perception is of about 7 photons. To our knowledge
this is the only firmly
established case in which a truly microscopic system can trigger directly a
definite perception. So,
let us start by considering the case in which a bunch, let us say of 10
photons coming from a region
A, propagates towards the eye of a human observer. The bunch hits the
retina of the observer and
triggers the definite perception ``a luminous spot at A''. Now we can
consider an analogous situation
in which the bunch comes from a region B, spatially separated and
perceptively distinct from A.
Again we can perform the experiment and the conscious observer will
perceive ``a luminous spot at
B''.

	Here comes the challenge for people working in quantum optics.
Suppose {\bf we are able to
 prepare} a state which is the superposition of the two previously
considered states, i.e.
the state\footnote{A state like the one we are considering has been
discussed for the
first time by Albert and Vaidman \cite{ghi15} (see also \cite{ghi16}) to
point out an
alleged difficulty of the dynamical reduction program. We have already
answered \cite{ghi17}
to this criticism proving its unappropriateness. In spite of that, the
analysis of these
authors has obviously played a precise role for the present paper. The
relevant fact,
however, is that at the time of the work by Albert and Vaidman the
possibility of
preparing such a state was a pure speculative assumption, while what we are
pointing out
in this paper is that now to prepare such a state seems practically
feasible.}:

\begin{equation}
|\Psi(1,2)\rangle\quad =\quad\frac{1}{\sqrt{2}}\left[| \mbox{10 photons
from A}\rangle
+ |\mbox{10 photons from B}\rangle\right]. \label{ghir4} \end{equation}
We also assume, {\bf and this is a crucial point of the proposal}, that
there is the practical possibility of
testing that actually the superposition persits for the time we will be
interested in. This means that
we have a mean to recombine the two states and to put into evidence their
interference effects, just
in a way analogous to the one we have analyzed above to test the fact that
the single photon is in a
superposition after traversing the birifrangent crystal. We do not know
whether the present
technology already allows such a performance, but Profs. De Martini and
Zeilinger have told me
that the desired result would certainly be attainable within a short time,
and that, for the time being,
to actually prepare it and to perform the interference test is mainly a
matter of money. Some
relevant steps in this direction have been performed recently \cite{ghi18,
ghi19}.

	Given these facts, we think that everybody will have clear the
novelty of the situation we are
envisaging with respect to the one analyzed in Section 2. Here we prepare a
state such that
we can {\bf test explicitly} that the quantum coherence between its terms
persists up to the moment in
which the conscious observers ``looks'' at it. The interesting fact is
that, even though the two
superposed states refer to a microscopic system, they can trigger definite
and different perceptions.
Here one cannot invoke wave packet reduction, the decoherence
induced by the entanglement with the environment, the possibility of making
the approximations
involved in the Bohmian analysis, or the decoherence induced by the
spontaneous localizations
{\it before} the observer looks at the photons. So we can proceed. We put
our observer in place of the
apparatus testing that the superposition is still there and we investigate
what are his perceptions.
We think everybody will agree about the probable outcome of the experiment:
the observer will
have definite perceptions and will not end up in a confused state of mind.
Moreover his perceptions
will be distributed randomly between ``I perceive a spot at A'' and ``I
perceive a spot at B''.

	But we also think that to perform the experiment is of some
relevance and that no one
would feel fully confident in answering to questions like: wich will be the
precise unfolding of the
perception process in this peculiar and fundamentally new situation? To
make clear our point of
view let us describe in greater details a precise set of experiments (among
the various ones) that can
be devised.

	We have an apparatus which can prepare three different kinds of
photon bunches:

\begin{enumerate}
\item[i)] Ten photons emanating from A,
\item[ii)] Ten photons emanating from B, \item[iii)] The above state
(\ref{ghir4}). \end{enumerate}

In the first run of experiments we stimulate the visual apparatus of the
observer by choosing
randomly the set-ups leading to the preparations i) or ii). Obviously the
observer will perceive a
random sequences of ``spots from A'' and ``spots from B''. In the second
run we stimulate the visual
apparatus of the observer by choosing always set-up iii). We espect that he
will still perceive a
random sequence of ``spots from A'' and ``spots from B''. But now we raise
the questions: are
 we sure that the specific details of the perception will be the same? For
instance, are
we sure that the reaction times and/or the possible perceptual failures
will be the same in the two cases? These questions are
legitimate and do not admit a naive answer (more about this in the next
Section) since one must
recall that, loosely speaking, in the second run the brain has to do the
``extra job''
 (with respect to the one of the first run) of ``reducing the
superposition'' of the initial
 nervous stimuli.

	These arguments should have made clear why we consider of some interest
to actually perform the experiment: it might be that we can get from it
some unespected
information, which could be of some relevance. At any rate we will at least
obtain the experimental
proof that a system like our sensory apparatus, contrary to the complicated
and macroscopic system
used to test that the superposition is still there, is actually able to
``reduce the statevector''.

	One could then think of more ambitious programs. For instance, the
late Prof. Borsellino
has repeatedly called my attention on the fact that even quite small living
organisms can be trained
to react to a light stimulus. One could then investigate the reactions of
such organisms to see
whether some differences occur when one subjects them to one of the two
above procedures. Many
variants of the indicated process can be devised. Obviously it is rather
hazardous to guess that
one could really learn something from such experiments since many loopholes
will in any case
remain open, but at the same time it seems to us that they have some
intrinsic interest.

\section{Making our argument more plausible.}

	To fully appreciate our point of view and the reasons for which we
believe that the proposed
experiment might be relevant it is appropriate to discuss a possible naive
objection against it. One
could argue along the following lines. Consider the first run of
experiments and the case in which
the preparation procedure leads to situation i). The photons impinge on the
observer's retina and
trigger the transmission of the nervous signal. Let us consider an
oversimplified description of the
unfolding of the process from the initial stimulus to an intermediate stage
in which part of the brain
is involved:

\begin{equation}
|\mbox{stimulus i); brain state R;...}\rangle \Rightarrow | \mbox{stimulus
i); brain state Si;...}\rangle. \label{ghir5} \end{equation}
Here R stays obviously for ``ready to react to the stimulus'' and we have
left unspecified (for reasons
which will become clear in a moment) the states of the other parts of the
perceptual apparatus
which will enter into play, in particular those referring to the final
stage in which a definite
perception emerge . Analogous considerations apply to the case in which the
initial state is the state
ii):

\begin{equation}
|\mbox{stimulus ii); brain state R;...}\rangle \Rightarrow | \mbox{stimulus
ii); brain state Sii;...}\rangle. \label{ghir6} \end{equation}

	One could then remark that since in each of the experiments of the
second run the initial
state is the superposition of the two above initial states, according to
the linear nature of quantum
evolution, one would have:

\begin{eqnarray}
& & \frac{1}{\sqrt{2}}\left[|\mbox{stimulus i); brain state R;...}\rangle +
|\mbox{stimulus ii); brain state R;...}\rangle\right] \nonumber \\ &
\Rightarrow & \frac{1}{\sqrt{2}}\left[|\mbox{stimulus i); brain state
Si;...} \rangle + |\mbox{stimulus ii); brain state Sii;...}\rangle\right].
\label{ghir7}
\end{eqnarray}
For such a state, however, the subsequent occurrence of wave packet
reduction (due e.g. to the
emergence of a definite conscious perception or to the uttering by the
observer of a sentence stating
which perception he has experienced, or to the coupling of his brain states
with the environment),
will lead precisely to one of the final states (\ref{ghir5}) or
(\ref{ghir6}). Thus, the proposed experiment would be
meaningless: at the end there is intrinsically no difference between the
two types of runs we have
proposed.

	However, it has to be stressed that the above argument is entirely
based on the assumption
that the linear and deterministic evolution characterizes all physical
processes and on the
hypothesis, characteristic of standard quantum mechanics, that the
reduction process is essentially
an instantaneous process. Moreover it is also assumed that such a process
simply selects one of the two terms of the superposition without in any way
affecting it. This is a
very restrictive and unrealistic assumption. Actually, the only request
that is necessary to yield
agreement with our definite experience is that the final state belongs to a
precise eigenmanifold
(among the two possible ones) of the
observable ``my perception about the location of the spot''. Now, the idea
that the assumptions listed
above are unappropriate is shared by an increasing number of scientists. We
can mention the lucid
statement by Leggett \cite{ghi20}: {\it one might imagine that there are
corrections to Schr\"{o}dinger's equation which are totally negligible at
the level of one, two, or even one hundred particles but play a major
role when the number of particles involved becomes macroscopic}, and the
repeated assertions of R.
Penrose \cite{ghi21} as emblematic examples. The very fact that the
dynamical reduction program has raised so much interest and has been so
convincingly supported by J.S. Bell, shows that the
scientific community is contemplating seriously the idea that one should
give up the unrestricted
validity of the linear nature of the evolution. If this is the case, the
physical processes which present
themselves as the most natural candidates to get some evidence of a
violation of the standard
dynamics are the mesoscopic processes and those processes like our definite
perceptions which
(even if we know very little about them) seem to require that reduction
actually takes place. It is
therefore reasonable to entertain the idea that in the case under
discussion a non linear and
stochastic dynamics, the one leading to wave packet reduction and to a
definite perception, governs
the process we are analyzing. Accordingly, there is no reason to exclude
that in spite of the fact that
the evolutions described by Eqs. (\ref{ghir5}) and (\ref{ghir6}) rule the
process when the initial states are i) or ii), respectively, the correct
equation which actually describes the unfolding of the process and which
has to replace Eq. (\ref{ghir7}) in case iii) will take the form:

\begin{eqnarray}
& & \frac{1}{\sqrt{2}}\left[|\mbox{stimulus i); brain state R;...}\rangle +
|\mbox{stimulus ii); brain state R;...}\rangle\right] \nonumber \\ &
\Rightarrow & \frac{1}{\sqrt{2}}\left[|\mbox{stimulus i); brain state
(Si)*;...}
\rangle + |\mbox{stimulus ii); brain state (Sii)*;...}\rangle\right].
\label{ghir8}
\end{eqnarray}
where the states (Si)* and (Sii)* are physically different (in some
respect) from the corresponding
states Si and Sii appearing in Eqs. (\ref{ghir5}) and (\ref{ghir6}).
Alternatively, one could contemplate
the possibility (see the next section for an illuminating example) that the
non linear evolution
might lead
to a final state like the one in Eq. (7), in which, howewer, a linear
superposition with
coefficients quite near but not exactly equal to $1/\sqrt{2}$, appears.
Then, at the moment in which
the subsequent evolution, e.g. the final act of perception or any other
physical mechanism,
will induce the
reduction, the actual state of the brain would be different according to
whether we are
performing
experiments of the first or of the second type, respectively. In turn,
these mesoscopic
differences
could give rise to slight differences concerning the reaction mechanisms of
the nervous
system,
such as, e.g., the average reaction times and/or possible perceptual failures.

	That the present argument has to be taken seriously, can be shown
by making reference to
one of the theoretical models which give a mathematically precise (even
though phenomenological)
account of the reduction process, i.e., the Spontaneous Localization Models
\cite{ghi9, ghi10, ghi11, ghi12}. Within such models the dynamical
processes leading to the final definite outcome and/or perception take place
during appropriate time intervals which are precisely defined by the
physical context. Actually, just
to answer the criticisms of ref. \cite{ghi15, ghi16}, we have analyzed
\cite{ghi17} (even though only in a qualitative way) precisely the visual
perception mechanism. What we have proved is that one can summarize
the unfolding of the process in the following way. As soon as the
superposition of the two stimuli
excite the retina, two nervous signals start and propagate along two
different axons. The process
goes on and the relative weights of the two terms of the superposition
change due to the stochastic
processes affecting the ions which cross the Ranvier nodes to transmit the
signal. Only when, in the
process of nervous transmission along the axons to the lateral geniculate
body and to the higher
visual cortex, the number of ions which occupy different positions in the
two states of the
superposition reaches an appropriate threshold, the suppression can be
considered complete: only
one of the two signals survives and triggers an unambiguous perception
process.
Has it has been shown in \cite{ghi17} the completion of the process
requires times comparable to
the perception time. This means that for an appreciable fraction of such a time
the superposition of the two "nervous signals" is still present, the two
terms having
comparable weights. In turn, each term will be associated to smeared
electric effects
involving few particles but characterized by a spreading of the
wavefunctions associated
to them. Then, if one
takes into account the very structure of the theory, one realizes that the
genuinely stochastic
processes implying the ``spontaneous localization'' of precise particles of
the system, are governed
by the overall wavefunction and, as a consequence, they might
exhibit  different features when one
consider {\it either} situation i) or situation ii) {\it versus} situation
iii).
 But a localization of a particle can
lead, as discussed in refs. \cite{ghi9, ghi10, ghi11, ghi12}, to the
excitation or dissociation of the ion to which it belongs. In brief, the
actual unfolding of the process in case iii) is by no means the ``linear
combination'' of the
processes occurring in the cases i) and ii). Thus, when the reduction will
actually come to an end
and the definite perception will emerge, it is not
unreasonable to admit that some precise and permanent record of
the fact that the perceptual apparatus has been triggered by case iii) and
not by {\it either} case i) {\it or} case ii), might remain. Whether such
differences can
have a systematic effect leading for instance to an increase of perceptual
failures
and errors or to a change in the reaction time is a fact that we cannot
give for granted but
which we cannot even exclude.

	Obviously, due to the fact that the considered dynamical reduction
models have only a
phenomenological status and that it has not be possible to test them
against quantum mechanics,
due to our extremely vague knowledge of percetual processes and so on, one
cannot take at its face
value the above argument and one cannot be precise about the differences we
have to look for. But
the general argument is surely correct: if some mechanism is present in
nature which actually leads
to reductions (and as such it must violate the linear nature of standard
quantum mechanics) then
this fact by itself implies that, after the reduction process has taken
place and has led to a definite
``outcome/perception'', the final state of all systems which have entered
into play in the process
 will be, in general, different  according whether the final perception has
been triggered by a
state which could lead to different perceptions or it has been triggered by
a state which can
only lead to the one which actually occurred. To clarify further this point
it seems usuful to
discuss in detail a dynamical reduction toy model which allows to focus the
specific features
of dynamical reduction theories in the case in which the reduction process
becomes competitive
with the hamiltonian evolution.

\section{A detailed study of a simple dynamical reduction model.}

In this section we will consider a dynamical reduction model of the type of
those introduced
in refs. [9-12]. However, since we intend to mimic the unfolding of a dynamical
 process in which a microscopic state triggers a process which takes an
appreciable time
(let us say of the order of the perceptual times) to lead to the situation
in which the nonlinear
and stochastic processes inducing the reduction become effective, we cannot
disregard (as one
 usually does
when discussing the suppression of superpositions of macroscopically
different states -
like ``$|${pointer here}$\rangle$+$|${pointer there}$\rangle$"), the
hamiltonian part of
the evolution equation. This can be easily understood: as discussed in
refs. [9-12], in
the case of superpositions corresponding to different locations of a
macroscopic object the
{\it effective} spontaneous localizations occurs with an extremely high
frequency $\lambda \cong 10^7
sec^{-1}$. Then, for a macroscopic system, the hamiltonian evolution during
two successive
localizations cannot change appreciably the state of the system and can be
disregarded. As a consequence
the state

\begin{equation}
|\Psi\rangle = a|\mbox{pointer here}\rangle
+ b |\mbox{pointer there}\rangle. \label{ghir9} \end{equation}
will be transformed, with probabilities exactly equal to $|a|^2$ and
$|b|^2$ into one
of the two terms of eq.(\ref{ghir9}).

The situation we are facing here is quite different: we are considering the
unfolding of a
physical process which at the beginning involves few particles, and
consequently no appreciable
contribution can derive from the nonlinear terms and the evolution is
essentially the hamiltonian one. Subsequently, the linear and nonlinear
dynamical terms
become competing up to the moment (after about $10^{-2} sec$) in which the
reducing terms govern
the game.
To get an idea of how things could go for the case we are interested in we
have to consider a
model
 in which the time between two successive reduction processes is such that,
during it, the
hamiltonian
evolution can modify  appreciably the state of the system. The detailed
discussion of a
situation of this type will allow
us to focus some extremely relevant features of dynamical reduction theories.

The most elementary model in which a competition between the hamiltonian
and the stochastic
and nonlinear
terms can be explicitly studied is the one of a system whose associated
Hilbert space is two-dimensional.
 Obviously, such a
toy model has nothing to do with the actual unfolding of a perceptual
process. But our aim is simply
to identify some precise formal features which characterize the evolution.
Actually, for our
limited
purposes, it will not even be necessary to discuss the dynamical behaviour
at the individual
level
(i.e. to follow the stochastic evolution of each individual physical
system), the relevant information
being deducible directly from  the statistical operator.

We will therefore consider a dynamics of the GRW or of the CSL type
(individual discontinuous or continuous reductions) and the evolution
equation for the statistical operator that it implies. Let us be more
precise about the details of the process. We have a system S in a two
dimensional Hilbert
space and we suppose that the linear part of its evolution is
governed by the Hamiltonian:
\begin{equation}
H=\hbar \omega \left|
\begin{array}{ll}
0 & A \\
A^{*} & 0
\end{array}
\right| ,\left| A\right| =1,  \label{Nw1}
\end{equation}
while the analogous of the spontaneous localizations are reductions on the
one-dimensional manifolds identified by the projection operators:
\begin{equation}
P_{+}=\left|
\begin{array}{ll}
1 & 0 \\
0 & 0
\end{array}
\right| ,P_{-}=\left|
\begin{array}{ll}
0 & 0 \\
0 & 1
\end{array}
\right| .  \label{Nw2}
\end{equation}
The reduction processes are assumed to take place with a mean frequency $%
\lambda $.

The dynamical evolution equation for the statistical operator is  of the
Quantum Dynamical Semigroup type and reads:
\begin{equation}
\frac{d\rho }{dt}=-\frac{i}{\hbar }\left[ H,\rho \right] +\lambda P_{+}\rho
P_{+}+\lambda P_{-}\rho P_{-}-\lambda \rho .  \label{Nw3}
\end{equation}
If we write

\begin{equation}
\rho (t)=\left|
\begin{array}{ll}
r(t) & \beta (t) \\
\beta ^{*}(t) & 1-r(t)
\end{array}
\right| ,  \label{Nw4}
\end{equation}
the evolution equations are:

\begin{eqnarray}
\frac{dr(t)}{dt} &=&2\omega \left[ \mathrm{Im}(A)\mathrm{Re}(\beta
(t))-\mathrm{Re}(A)%
\mathrm{Im}(\beta (t))\right]  \label{Nw5} \\
\frac{d\mathrm{Re}(\beta (t))}{dt} &=&\omega \mathrm{Im}\left[ A\left\{
1-2r(t)\right\} \right] -\lambda \mathrm{Re}(\beta (t))  \nonumber \\
\frac{d\mathrm{Im}(\beta (t))}{dt} &=&-\omega \mathrm{Re}\left[ A\left\{
1-2r(t)\right\} \right] -\lambda \mathrm{Im}(\beta (t)).  \nonumber
\end{eqnarray}

A general theorem by Spohn \cite{Spohn} guarantees that, in the finite
dimensional case, if a Quantum Dynamical Semigroup equation admits a steady
solution, then the general solution converges to it for $t\rightarrow
+\infty .$ In our case the steady solution is $\rho (t)=\frac{1}{2}I.$
Taking into account this fact one easily writes the general solution of our
problem:

\begin{eqnarray}
r(t) &=&\left\{ \frac{2\omega }{\lambda \Delta }\left[
\mathrm{Re}(A)\mathrm{Im}%
(\beta (0))-\mathrm{Im}(A)\mathrm{Re}(\beta (0))\right] +\frac{4\omega
^{2}\left[
1-2r(0)\right] }{\lambda ^{2}\Delta \left[ 1-\Delta \right] }\right\} e^{-%
\frac{\lambda }{2}(1+\Delta )t}  \nonumber \\
&&-\left\{ \frac{2\omega }{\lambda \Delta }\left[
\mathrm{Re}(A)\mathrm{Im}(\beta
(0))-\mathrm{Im}(A)\mathrm{Re}(\beta (0))\right] +\frac{4\omega ^{2}\left[
1-2r(0)\right] }{\lambda ^{2}\Delta \left[ 1-\Delta \right] }\right\} e^{-%
\frac{\lambda }{2}(1-\Delta )t}+\frac{1}{2}  \nonumber \\
\mathrm{Re}\beta (t) &=&a\mathrm{Im}(A)e^{-\frac{\lambda }{2}(1+\Delta
)t}+b\mathrm{%
Im}(A)e^{-\frac{\lambda }{2}(1-\Delta )t} \nonumber\\
&&+\left[ \mathrm{Re}(A)\mathrm{Re}(\beta (0))+%
\mathrm{Im}(A)\mathrm{Im}(\beta (0))\right] \mathrm{Re}(A)e^{-\lambda t}
\label{New6} \\
\mathrm{Im}(\beta (t)) &=&\left[ \mathrm{Re}(A)\mathrm{Re}(\beta
(0))+\mathrm{Im}(A)\mathrm{Im}%
(\beta (0))\right] \mathrm{Im}(A)e^{-\lambda t}\nonumber\\
&&-a\mathrm{Re}(A)e^{-\frac{\lambda }{2}
(1+\Delta )t}-b\mathrm{Re}(A)e^{-\frac{\lambda }{2}(1-\Delta )t}.  \nonumber
\end{eqnarray}
In the above equation we have put:
\begin{equation}
\Delta =\sqrt{1-16\varepsilon ^{2}},\varepsilon =\frac{\omega }{\lambda }.
\label{New7}
\end{equation}

We have now to choose precise values for $\omega $ and $\lambda .$ Let us
suppose that $\varepsilon $ is smaller than $\frac{1}{4}.$ From eqs. (\ref
{New6}) we see that $\rho (t)\rightarrow \frac{1}{2}I$ for $t\rightarrow
+\infty ,$ independently from its initial value $\rho (0).$ This, in
particular, means that for {\it extremely long times} the reducing dynamics
will not work as one would like, i.e., it will not transform the pure state $%
\left| \Psi \right\rangle =a\left(
\begin{array}{l}
1 \\
0
\end{array}
\right) +b\left(
\begin{array}{l}
0 \\
1
\end{array}
\right) $ into either $\left(
\begin{array}{l}
1 \\
0
\end{array}
\right) $ or $\left(
\begin{array}{l}
0 \\
1
\end{array}
\right) $ with the desired probabilities $\left| a\right| ^{2}$ and $\left|
b\right| ^{2},$ respectively \footnote{Actually, in the realistic case of
the GRW and
CSL models the situation is even worse since $\rho (t)$ does not have a
precise limit for $t\rightarrow +\infty $ (see, however, the comments
following
eq. (21)).}.

Suppose now that $\varepsilon $ is quite small. Then, within the time interval
 $\lambda
^{-1}<<t<<\left( 4\lambda \varepsilon ^{2}\right) ^{-1}$ we have:
\begin{equation}
e^{-\lambda t}\simeq 0,e^{-4\lambda\varepsilon ^{2}t}\simeq 1,  \label{New8}
\end{equation}
and the statistical operator
is practically diagonal with diagonal elements
corresponding (almost exactly) to the probability of reduction on the desired
manifolds. In fact, evaluating, for the  considered time interval  and at
first
order in
the small parameter $\varepsilon ,$ the explicit form of the element $r(t)$
of $\rho (t)$ for two different initial conditions, i.e.:
\begin{eqnarray}
\rho _{Mixt}(0) &=&\left(
\begin{array}{ll}
\left| a\right| ^{2} & 0 \\
0 & \left| b\right| ^{2}
\end{array}
\right) ,  \label{New 9} \\
\rho _{Pure}(0) &=&\left(
\begin{array}{ll}
\left| a\right| ^{2} & ab^{*} \\
a^{*}b & \left| b\right| ^{2}
\end{array}
\right) ,  \nonumber
\end{eqnarray}
corresponding, respectively, to the statistical mixture
\begin{equation}
\rho _{Mixt}(0)=\left| a\right| ^{2}P_{+}+\left| b\right| ^{2}P_{-},
\label{New10}
\end{equation}
and to the pure state
\begin{equation}
\rho _{Pure}(0)=\left| \Psi \right\rangle \left\langle \Psi \right| ,\left|
\Psi \right\rangle =a\left(
\begin{array}{l}
1 \\
0
\end{array}
\right) +b\left(
\begin{array}{l}
0 \\
1
\end{array}
\right) ,  \label{New11}
\end{equation}
we have:
\begin{equation}
r_{Mixt}(t)=\left| a\right| ^{2},r_{Pure}(t)=\left| a\right|
^{2}-2\varepsilon \left[ \mathrm{Re}(A)\mathrm{Im}({\it
ab}^{*})-\mathrm{Im}(A)\mathrm{Re}
({\it ab}^{*})\right] .  \label{New 12}
\end{equation}
In the considered approximation the off-diagonal elements of $\rho (t)$ are
equal to zero in the first case and of order $\varepsilon $ in the second.

Eqs.(\ref{New8}) and (\ref{New 12}) teach us a quite interesting lesson:

i. The reduction regime can last only for a certain (possibly extremely
long) time interval,

ii. The dynamical reduction process can lead, in such a time interval, to
different situations
 according whether the initial condition corresponds
to a statistical mixture of two states with weights  $|a|^2$ and $|b|^2$ or
to the  superposition
with coefficients {\it a} and {\it b} of the same states.

Obviously, the two above facts do not change in any way the relevant
implications of
the GRW and CSL theories. In fact,
such theories lead to the objectification of the positions of macroscopic
bodies, and, more generally, to a definite  \cite{ghi12}  average mass density
distribution (the average being evaluated on cells of the order of
$10^{-15}cm^{3}$ around any chosen
space point {\bf r }). If one would establish a correspondence
between the dynamics of these theories and the one of the toy model we have
analized in this section, one should take into account that the average
effective localization frequency for a macroscopic object within the GRW and
CLSL models corresponds to the choice $\lambda \simeq 10^{7}\sec ^{-1}$.
Moreover,
the hamiltonian cannot induce, during the time interval between two
reductions, appreciable transitions between the different macrostates to
 which the reduction mechanism leads, so that one can assume $\omega \simeq
0.$ Concluding,
in the macroscopic case
the dynamical reduction models
lead to a definite macroscopic situation in extremely short times
and such a situation persists (in absence of other dynamical processes)
practically forever.

Let us now  analyze the case  in which
the hamiltonian evolution and the reduction mechanism are competitive. Since
we want to mimic a perception process we could tentatively make in the above
toy model the following choice for the parameters characterizing it:

\begin{equation}
\lambda \simeq 10^{2}\sec ^{-1},\varepsilon \simeq 10^{-4}.  \label{New13}
\end{equation}
The first choice corresponds to the idea that the time which is necessary
for a definite perception to emerge is of the order of one hundreth of a
second, the second is done in order to guarantee that the statistical
distribution of the reduced states is (almost) perfectly respected, i.e.,
that the asymptotic regime leading to the equal weight distribution of the
two outcomes (independently from the triggering state), cannot become
immediately effective. That this is the case follows from the fact that the
above choice
implies  $4\lambda \varepsilon ^{2}\simeq 10^{-6}\sec ^{-1}.$

We can now
discuss some particular cases. Suppose one chooses in eqs. (\ref{New 9}), $%
a=b=\frac{1}{\sqrt{2}}$ , and in eq. (10), $A=i.$ Then eqs. (\ref{New 12})
become:
\begin{equation}
r_{Mixt}(t)=\frac{1}{2},r_{Pure}(t)=\frac{1}{2}-\varepsilon \simeq \frac{1}{2%
}-10^{-4}.  \label{New14}
\end{equation}
If we take this result as giving some indications about the perception
processes
 we have analyzed in section 5, we are led to conclude that when
we trigger the perceptual apparatus with the exact 50\%-50\% mixture of the
states $|$stimulus i)$>$ and $|$stimulus ii)$>$, the observer perceives
precisely as many ''spots at A'' as ''spots at B''. On the contrary, when
triggered by
the superposition $\frac{1}{\sqrt{2}}\left[ \left| {\rm stimulus}%
\,i)\right\rangle +\left| {\rm stimulus}\,ii)\right\rangle \right] $, he
will tend to perceive a slightly larger number of ''spots at B'' than ''spots
at A''.

Two remarks are at order:

i: We do not intend to suggest that the toy model has to be taken as
appropriately mimicking  the perceptual process. Our analysis has to be
considered as a rigorous proof that when the reduction process takes some time
to be completed and is competing with other processes, then, within the
only known consistent
theory describing such a process, some precise differences between
the outcomes generated by an initial statistical mixture and a pure state
(for which the squares of the moduli of its coefficients coincide with the
weights of the statistical mixture) can, and, in general will, actually
emerge.

ii. We stress that if one goes through the same calculation by changing the
sign of the second term of
the superposition we  have chosen as the initial state (i.e. if
consideration is given to
the state $\frac{%
1}{\sqrt{2}}\left[ \left| {\rm stimulus}\,i)\right\rangle -\left| {\rm %
stimulus}\,ii)\right\rangle \right] )$, then the expression for $r(t),$ in
the considered approximation, is $r(t)\simeq \frac{1}{2}%
+\varepsilon .$ The change of  sign in front of $\varepsilon $
 is extremely important. If the corrections to $\frac{1}{2}$
would not cancel when the + or - signs are considered, one could use the
difference in the
probabilities of the two outcomes  to set up a device allowing
faster-than-light signalling. This
cannot happen, and actually there are many proofs (the first one having been
exhibited by J. Bell \cite{Bell2}) that the GRW\thinspace theory (as well
as all models
involving dynamical reductions of the type of CSL) cannot lead to
superluminal communication.

Concluding: we have analyzed in detail a
toy model of dynamical reduction in which the
hamiltonian dynamics and the reduction mechanism compete and we have proved
that it
 allows to distinguish between an initial state corresponding to a linear
superposition and
 a statistical mixture
with the  appropriate weights. Obviously, since we have dealt with a two
dimensional Hilbert space, the difference could only consists in a
slight discrepancy of the probabilities of the two outcomes. It is trivial
(even
though a little bit tedious) to show that when analogous dynamical
conditions (the competing of the linear and nonlinear parts of the dynamics)
occur in a system whose Hilbert space is higher-dimensional and in which
the reductions occur
on degenerate manifolds, the differences
originating from different initial situations could also derive from the
different composition of the final ensemble within the manifolds to which
 reductions  lead. We have worked out  detailed calculations
for a four-dimensional case with reduction on two orthogonal
two-dimensional subspaces: we do not reproduce our results here since they
 involve quite cumbersome formulae. Those who have
followed our argument can easily grasp that things actually go as indicated
above.

The analysis of this section makes, in our opinion, quite
plausible the general argument developed in  the previous sections.

\section{Concluding remarks.}

We think that the above analysis, even though it gives only some general
hints about what might emerge in the experiments we have proposed, shows that
it is worth to actually perform them. The proposal identifies new ways
(which only recent technological developements have made possible) of
investigating such fundamental processes like the violation of the linear
nature of quantum mechanics and the formation of definite perceptions. As
already mentioned, R. Penrose \cite{ghi21}, as well as many other
scientists, have repeatedly suggested that strict links should exists
between the possible violation of the linear Schr\"{o}dinger evolution and
the perception mechanism. The present proposal points out that there is a
possibility of performing experiments which might throw some light precisely
on the crucial problems we have mentioned.

\section*{Acknowledgements.}

We thank Profs. F. De Martini and A. Zeilinger for exchanges of ideas on
this problem and for having informed us that the experiments we suggest seem
not far from being actually feasible. Their remarks have encouraged us to
write this ``divertissment''. We also acknowledge useful discussions with
Drs. A. Bassi and F. Benatti. A particular acknowledgement goes to Prof. C.
Bennet. When I have presented the argument of this paper at the last
Conference on Quantum Interferometry held in Trieste he immediately remarked
that the hypothesized mechanism which allows to distinguish, after the
perceptual process, a statistical mixture from a pure state would imply the
possibility of faster-than-light communication. I replied that this could
not be since general theorems show that dynamical reduction
mechanisms do not allow such a possibility. However, Prof. Bennet's remark has
pushed me to reconsider the alleged possibility, which is at the very basis
of the argument, that the dynamical reduction mechanism could actually work
as I
had assumed (without any detailed proof and resorting only to plausibility
arguments). Thus, in spite of the fact that a previous version
of the present paper had already been accepted for publication, I decided
that a more precise investigation of the reduction mechanism under conditions
like those I was envisaging was necessary. In particular one had to
 prove (and not simply to hypotesize) that the alleged differences can
actually occur, at least
under appropriate circumstances. The present version of this paper differs
from the
previous one only for the addition of section 6, whose interest derives from
 the explicit proof that what I had assumed can actually occur for an
extremely reasonable reduction model and from having made clear that,
 in spite of this, no faster-than-light signalling can
be achieved. This enrichment of the paper is entirely due to the cute remarks
by Prof. Bennet.

\end{document}